# Quality assurance for the ALICE Monte Carlo procedure

[1]Muhammad Ajaz and [2]Seforo Mohlalisi

(Summer Students CERN 2008)

Superviser: Peter Hristov (CERN, Switzerland)

Co Supervisor: Jean-Pierre Revol (CERN, Switzerland)

[1]Physics department, COMSATS IIT, Islamabad, Pakistan.

[2]Physics department, University of Cape town, South Africa.

1. **ALICE experiment:**

The aim of the ALICE experiment [1] is to study the physics of strongly interacting matter under extreme condition of high temperature and baryon densities, where the formation of a new phase of strongly interacting matter, the quark-gluon plasma, is expected. The ALICE detector is made up of the central detector and the forward muon arm. The central detector is embedded in a large solenoidal magnet with a weak field of 0.5 tesla, parallel to z, and it consists of the Inner Tracking System with six layers of high resolution silicon detectors, the Time Projection Chamber, which provide track finding, charged particle momentum measurement, particle identification, and two-track separation in the region $p_t$<10 GeV/c and pseudo-rapidity $|\eta|$<0.9, a Transition Radiation Detector for electron identification, a barrel Time of Flight dedicated to charged particle identification, a small area ring imaging Cherenkov detector at large distance for the identification of high momentum particles, and a single arm electromagnetic calorimeter. Outside the central barrel is the muon spectrometer, which is designed to measure the production of complete spectrum of heavy quark resonances, a forward photon counting detector (PMD), a multiplicity detector covering the forward rapidity region (FMD), A system of scintillators (V0 detector) and quartz counters (T0 detector) provide fast trigger signals.

One of the distinctive features of ALICE is the particle identification capability, which is realized using a number of different techniques. Charged hadron identification is provided over the full barrel acceptance ($|\eta|$ < 0.9) by the combination of (a) dE/dx measurement in the four outer layers of the ITS and in the TPC, for momenta up to ≈ 0.5 GeV/c, with (b) a barrel Time of

Flight, in the range 0.5 < p < 2.5 GeV/c. Electron are separated from pions for pt > 1 GeV/c by means of a dedicated Transition Radiation Detector and by exploiting the relativistic rise of the specific energy loss measured in the TPC. A smaller-area ring imaging Cherenkov detector (HMPID), covering about 15% of the acceptance of the ALICE central detectors, allows the separation of hadrons up to higher momenta (π/K up to 3 GeV/c and K/p up to 5 GeV/c). Photons and neutral pions are identified in the small-acceptance electromagnetic calorimeter PHOS [2].

After hadronization, particles emanating from the interaction vertex travel across the ALICE detectors leaving deposits of charge in detectors. In order to identify which particles were produced from the collision, the tracks the particles made as they moved across the detectors are reconstructed.

## 2. Offline Reconstruction:

The reconstruction [3] is done by the reconstruction frame work in the ALICE off-line framework, ALIROOT. The inputs to the reconstruction frame work are the digits. These are the digitized signals obtained by the sensitive pads of the detectors at a certain time. Digits can be in the .root files or they can be generated from the raw data (DDL). The intermediate output of the reconstruction is the cluster and is defined as the set of adjacent digits, in space and time, presumably generated by the same particle as it traversed the detectors. During clusterazation stage the reconstruction is done for each detector separately with no information exchanged between detectors, hence clusterazation is purely a local reconstruction.

The reconstruction frame work then calculates the center of gravity of the cluster to estimate the actual space point where the particle crossed the sensitive part of the detector and the interaction vertex is reconstructed from the data provided by the silicon pixel detectors of the ITS. The space points can be joined to construct the sub tracks and which can be joined to make a track. The track is extrapolated to the primary vertex. Tracks which do not originate from the primary vertex are discarded as background. The algorithm used for the offline reconstruction of the track is the kalman filtering and it takes into account multiply scattering and energy loss as it is reconstructing the tracks.

The reconstructed tracks, the interaction vertex and all data that is useful for physics analysis is stored in the Event Summary Data (ESD) file. There are two types of analysis that can be performed on data and these are scheduled and chaotic analysis.

### 3. Analysis Framework:

The analysis is done using analysis tasks. The task is implemented as a derived class that inherits from the base class AliAnalysisTaskSE. However the full implementation is needed for a task then it should inherit directly from the base class AliAnalysisTask.
The mandatory methods that have to be overridden are UserCreatOutputObjects, UserExec and Terminate. In the method UserCreateOutputObjects all objects, histograms, trees, etc that will represent the output of the analysis module have to be initialized. The analysis algorithm to be executed per event are called or implemented in UserExec and the histograms are filled here. The access to the event is provided by the Monte Carlo event handler. Finally the histograms represent the output data are drawn in the method Terminate.

Analysis Task can be run locally on the personal computer, on the CERN analysis framework (CAF) or on the grid. In all of these cases the analysis train has to be implemented in the form of a ROOT macro. The analysis train is the way to run analysis in the most efficient way over a large part or the full dataset. It is using the AliAnalysisManager framework to optimize CPU/IO ratio, accessing data via a common interface and making use of PROOF and GRID infrastructures.

The train is assembled from a list of modules that are sequentially executed by the common *AliAnalysisManager* object. All tasks will process the same dataset of input events, share the same event loop and possibly extend the same output AOD with their own information produced in the event loop.

The first thing to do when writing the chain macro to run on proof is to open the connection to the proof cluster with the statement TProof::Open("username@lxb6046"). Then the managing packages are uploaded with statements gProof->UploadPackage("package version") and gProof->EnablePackage("package version"). After uploading the managing packages the analysis manager and the task are created and the task is added to the manager. There should at least be one task per analysis chain. The ESD handler object is then created to access the ESD and the TChain to collect input files is also created. The containers for the inputs and for the outputs are created and connected to the slots by the manager. Futhermore checks are performed for data type consistency and any illegal circular dependency between the modules is signaled. Then the analysis is started in the proof mode. The detailed macro that implements the analysis train is given in appendix A[4].

### 4. Quality assurance for MC productions:

Our task was to implement the already existing macro, $ALICE_ROOT/STEER /CheckESD.C that is ran after reconstruction to compute the physics efficiency, as a task that will run on proof framework like CAF. The task was implemented in a C++ class called AliAnalysisTaskCheckESD and it inherits from AliAnalysisTaskSE base class. The function of AliAnalysisTaskCheckESD is to compute the ratio of the number of reconstructed particles to the number of particle generated by the Monte Carlo generator. The AliAnalysisTaskCheckESD does not either enhance or reduce the functionality of the CheckESD.C macro but rewrites it as an analysis task. Thus the parts of CheckESD.C that initializes the histograms are put in the method AliAnalysisTaskCheckESD:: UseCreateOutputObjects(), the parts that that loop over the event and hence implement the analysis algorithm for the event are put in AliAnalysisTaskCheckESD::UserExec(), and the parts that draw the histograms are put in the AliAnalysisTaskCheckESD::Terminate() method.

The first thing that the program make sure that the particles that we deals with have the transverse momentum greater than 0.001GeV and its pseudorapidity is with in ±0.9. If the particle is a proton with transverse momentum ($P_t$) is greater than 0.1GeV or if it is a $\Lambda_0$ with $P_t$ greater than 0.3GeV or it's a Ω⁻ with $P_t$ greater than 0.5GeV then it is selected for further processing. On this selected set the tracks which are background are removed.

Then we loop over the muon tracks and select the ones whose inverse bending momentum is greater than 0.001 GeV. Similar loops over the V0, the cascades, the clusters are implemented to select true particles over background and the corresponding histograms are filled. At the end of the AliAnalysisTaskCheckESD::UserExec() method, the histograms are posted so that they can be retrieved in other methods like Terminate for further processing.

In the method AliAnalysisTaskCheckESD::Terminate, the efficiency is calculated as the ratio of the number of reconstructed tracks to the number of generated tracks.

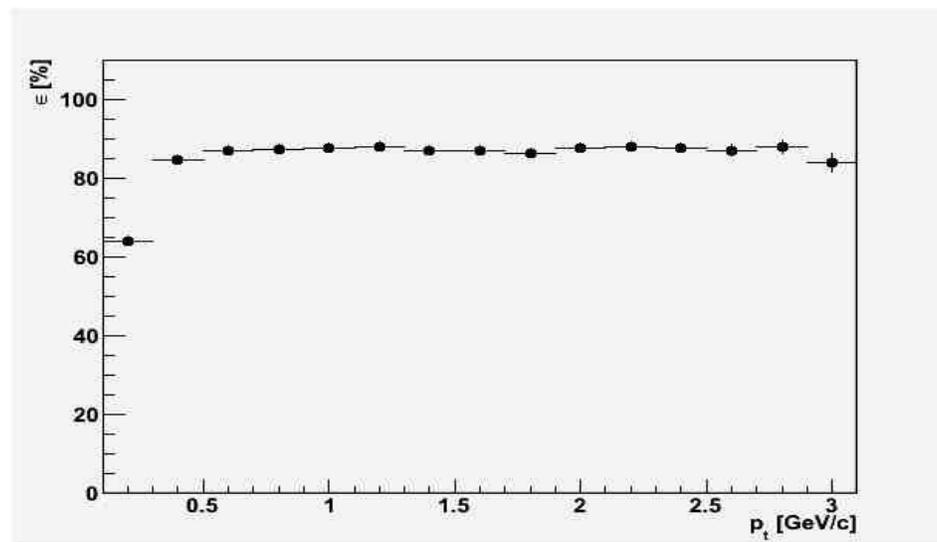

**Graph 1. The efficiency histogram.**

The ratio of the fake tracks to the generated tracks is also calculated. The resolution in Phi, $P_t$ and theta are calculated.

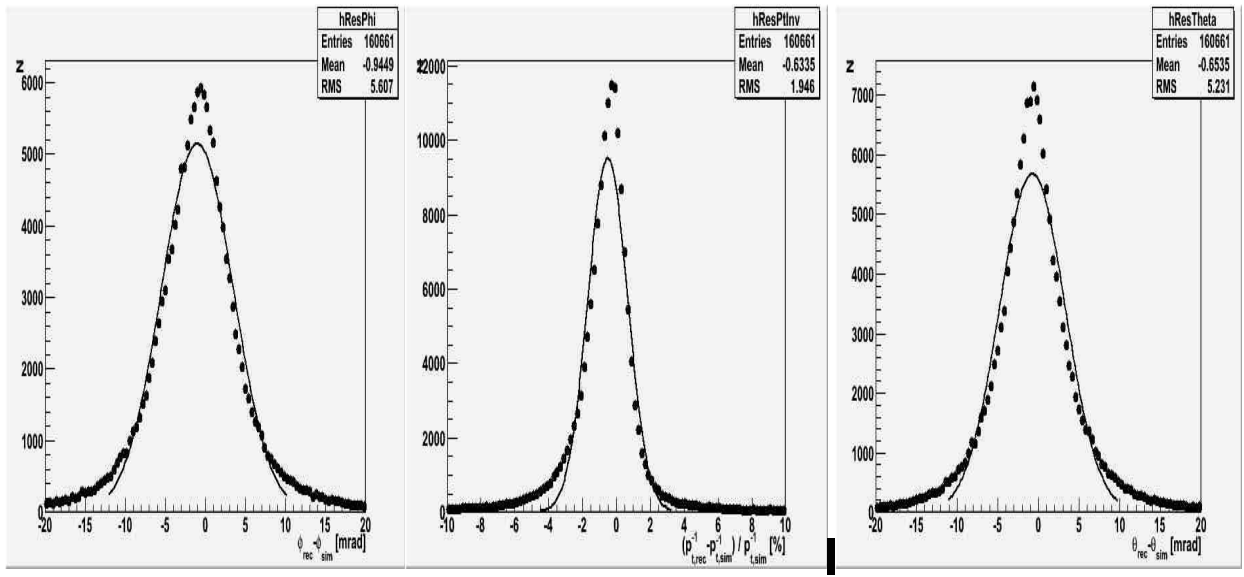

**Graph 2. Phi, $P_t$ and theta resolution.**

Then the particle Identification efficiency is calculated by the ratio of number of identified particles to the number of reconstructed tracks.

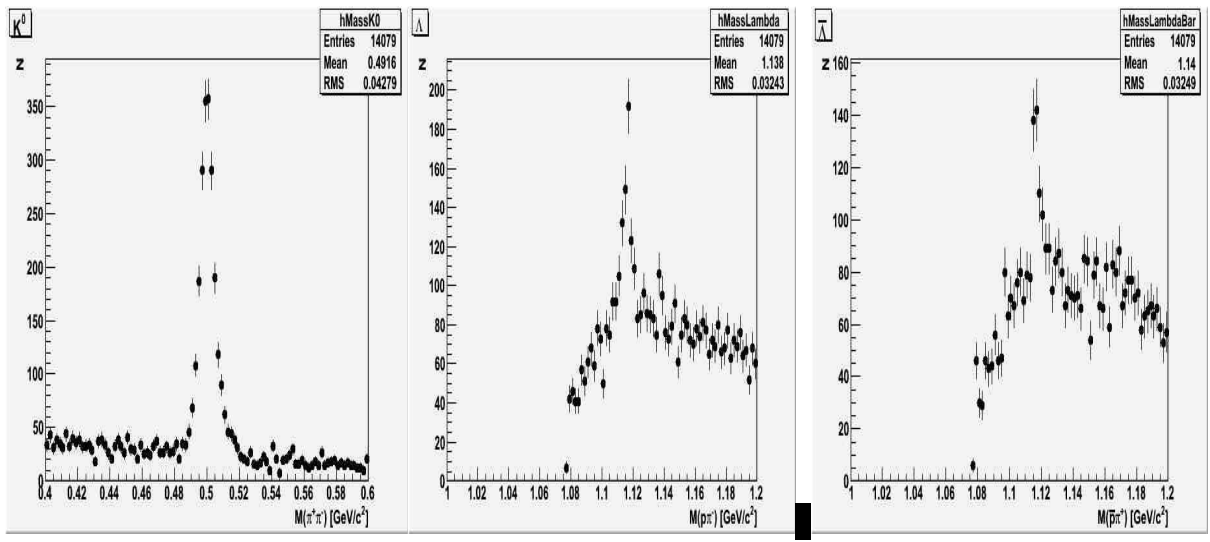

**Graph 3. K$^0$, Λ and Λ$^-$**

The header file of the class AliAnalysisTaskCheckESD class is appendix B[5] and the source file is in appendix C[6].

## 5. Conclusion:

The class AliAnalysisTaskCheckESD was successfully implemented. It was used during the production for first physics and permitted to discover several problems (missing track in the MUON arm reconstruction, low efficiency in the PHOS detector etc.). The code is committed to the SVN repository and will become standard tool for quality assurance.